\documentclass[10pt,conference]{IEEEtran}
\IEEEoverridecommandlockouts
\usepackage{cite}
\usepackage{amsmath,amssymb,amsfonts}
\usepackage{algorithmic}
\usepackage{graphicx}
\usepackage{textcomp}
\usepackage{xcolor}
\usepackage{xspace}
\usepackage{fontawesome}
\usepackage{multirow}
\usepackage{rotating}
\usepackage{tikz}
\usepackage{soul}
\usepackage{makecell}
\usepackage{url}
\usepackage{enumitem}
\usepackage{booktabs}
\usepackage{fontawesome}
\usepackage{pifont}
\usepackage{amssymb}
\usepackage[colorlinks=true, allcolors=black]{hyperref}

\newcommand{\cmark}{\ding{51}}%
\newcommand{\xmark}{\ding{55}}
\newcommand{\ie}{\emph{i.e.,}\xspace}
\newcommand{\eg}{\emph{e.g.,}\xspace}
\newcommand{\etc}{etc.\xspace}
\newcommand{\etal}{\emph{et~al.}\xspace}
\newcommand{\secref}[1]{Section~\ref{#1}\xspace}

\newcommand{\figref}[1]{Fig.~\ref{#1}\xspace}

\newcommand{\tabref}[1]{Table~\ref{#1}\xspace}

\newcommand{\relevant}{33\xspace}
\newcommand{\pretrained}{32\xspace}
\newcommand{\hours}{890\xspace}

\def\BibTeX{{\rm B\kern-.05em{\sc i\kern-.025em b}\kern-.08em
    T\kern-.1667em\lower.7ex\hbox{E}\kern-.125emX}}

\begin{document}

\title{Automating Code-Related Tasks Through Transformers: The Impact of Pre-training}

\author{
\IEEEauthorblockN{Rosalia Tufano, Luca Pascarella, Gabriele Bavota}
\IEEEauthorblockA{\textit{Software Institute -- USI Universit\`{a} della Svizzera italiana, Switzerland}}
}

\maketitle

\begin{abstract}
Transformers have gained popularity in the software engineering (SE) literature. These deep learning models are usually pre-trained through a self-supervised objective, meant to provide the model with basic knowledge about a language of interest (\eg Java). A classic pre-training objective is the \emph{masked language model} (MLM), in which a percentage of tokens from the input (\eg a Java method) is masked, with the model in charge of predicting them. Once pre-trained, the model is then fine-tuned to support the specific downstream task of interest (\eg code summarization). While there is evidence suggesting the boost in performance provided by pre-training, little is known about the impact of the specific pre-training objective(s) used. Indeed, MLM is just one of the possible pre-training objectives and recent work from the natural language processing field suggest that pre-training objectives tailored for the specific downstream task of interest may substantially boost the model's performance. For example, in the case of code summarization, a tailored pre-training objective could be the identification of an appropriate name for a given method, considering the method name to generate as an \emph{extreme summary}. In this study, we focus on the impact of pre-training objectives on the performance of transformers when automating code-related tasks. We start with a systematic literature review aimed at identifying the pre-training objectives used in SE. Then, we pre-train \pretrained transformers using both (i) generic pre-training objectives usually adopted in SE; and (ii) pre-training objectives tailored to specific code-related tasks subject of our experimentation, namely bug-fixing, code summarization, and code completion. We also compare the pre-trained models with non pre-trained ones and show the advantage brought by pre-training in different scenarios, in which more or less fine-tuning data are available. Our results show that: (i) pre-training helps in boosting performance only if the amount of fine-tuning data available is small; (ii) the MLM objective is usually sufficient to maximize the prediction performance of the model, even when comparing it with  pre-training objectives specialized for the downstream task at hand.
\end{abstract}


\section{Introduction} \label{sec:intro}

Transformers \cite{vaswani2017attention} are deep learning (DL) models built on the idea of self-attention, a mechanism able to assign different weights to different parts of the input data. Transformers achieved state-of-the-art performance in several natural language processing (NLP) tasks and, recently, have gained popularity in the software engineering (SE) literature for the automation of code-related tasks (see \eg \cite{Drain:maps2021,Berabi:pmlr,Mastropaolo:icse2021,Mashhadi:msr2021,Yang:icpc2021,Peng:2021,Tang:2021,Tufano:icse2022,Patanamon:icse2022,Li:fse2022,Svyatkovskiy:fse2020,Ciniselli:tse2021}). 

Transformers are usually trained in a two-step process. First, they are pre-trained using a self-supervised training objective with the goal of providing the model with general knowledge about a language relevant for the final task to automate. Then, the model is fine-tuned to support the downstream task of interest. 

The most adopted pre-training objective is the \emph{Masked Language Model} (\textsc{mlm}), in which a percentage of tokens from the input sentence is masked, with the model in charge of predicting them. For example, given the final goal of building a translator from English to German, a transformer can be firstly pre-trained by feeding as input English and German sentences with masked words. This task is self-supervised, since random words can be automatically masked with the model in charge of predicting them. Then, a labeled dataset mapping English sentences to their corresponding German translation can be used to fine-tune the model.
Several works applying DL to SE report boost of performance\footnote{With ``performance'' we refer to the quality of the predictions generated by the model (\eg accuracy) rather than to attributes such as execution time.} provided by pre-training in the automation of code-related tasks \cite{Li:fse2022,Ciniselli:tse2021,Mastropaolo:tse2022}. However, little is known about (i) the circumstances in which pre-training actually helps, and (ii) the impact of the specific pre-training objective(s) adopted on the performance of transformers when automating code-related tasks. 

Concerning the first point, it is known that pre-training is helpful when the fine-tuning dataset is small \cite{Robbes:icse2019}. To make a concrete example, per-training can be useful when fine-tuning a model for automated bug-fixing. For this task, the fine-tuning dataset is usually composed by pairs of $\langle$\emph{buggy}, \emph{fixed}$\rangle$ functions, which are mined from bug-fixing commits. The amount of training data that can be collected is thus limited, in the order of tens of thousands instances \cite{Tufano:tosem2019,Chen:tse2021}. Pre-training datasets, instead, can be automatically built with virtually ``no limitations'' in terms of size since the pre-training objective is self-supervised. There are however SE tasks which can leverage very large fine-tuning datasets, for which the boost in performance provided by pre-training is not obvious. For example, in the case of code completion, fine-tuning instances are usually represented by pairs of $\langle$\emph{incomplete\_code}, \emph{complete\_code}$\rangle$, with the model learning how to ``finalize an implementation task''. These pairs can be built automatically from any piece of code by removing parts of it.

As for the second point (\ie the impact of pre-training objectives), the widely used \textsc{mlm} is just one of the possibilities here: Any self-supervised task can be used as pre-training objective. Also, in a recent work from NLP, Zhang \etal hypothesized that ``\emph{using a pre-training objective that more closely resembles the downstream task leads to better fine-tuning performance}'' \cite{Zhang:ICLM2020}. 

\eject
 
We present a large-scale study aimed at (i) investigating whether pre-training is actually useful in code-related tasks for which the fine-tuning dataset can be built without any obvious impediment in collecting large amount of data (\eg code completion); and (ii) experimenting the impact on transformers' performance of both generic and task-specific pre-training objectives when automating three code-related tasks, namely bug-fixing, code summarization, and code completion. 

We start by performing a systematic literature review (SLR) to identify the pre-training objectives used in the SE literature. We inspected a total of 936 papers published in SE venues up to January 2022, ending up with \relevant relevant papers, namely those (i) automating a code-related task (\ie a task involving source code as input and/or output of the model), and (ii) pre-training a transformer or exploiting an already pre-trained transformer from previous work. The output of such a SLR (\tabref{tab:objectives}) provided us with an overview of pre-training objectives used in the literature.

Based on this analysis, we selected three generic pre-training objectives to experiment based on their popularity and potential impact in SE: \textsc{mlm}, \emph{next sentence prediction}, and \emph{replaced tokens detection}. Moreover, we defined three pre-training objectives tailored for the specific downstream tasks we aim at supporting. For \emph{bug-fixing}, we pre-train the model through the \emph{injected mutants fixing} objective. The latter simulates the bug-fixing downstream task (i) without using real bug-fixing instances that can be kept to increase as much as possible the size of the fine-tuning dataset; and (ii) while being scalable in terms of size, since the objective is self-supervised and the experimenter can create as many pre-training instances as wished. Concerning the \emph{code summarization} task, we consider the \emph{Method Name Generation} as a tailored self-supervised pre-training objective: The model takes as input a method and must guess an appropriate name for it. The name is treated as an \emph{extreme summary} \cite{allamanis2016convolutional}, thus preparing the model for the downstream task in which, given a method, it is expected to generate a textual summary. Finally, concerning \emph{code completion}, we devise the \emph{code block selection} objective, which asks the model to guess which among two code blocks is appropriate to complete a code snippet having a code block masked. Such an objective prepares the model for the more challenging downstream task, in which it is asked to generate from scratch the missing block rather than just selecting it among two choices.

We then trained 36 Text-To-Text Transfer Transformer (T5) models \cite{raffel2019exploring}, accounting for a total of \hours training hours, using different pre-training objectives and fine-tuning datasets. In particular, we study the impact of the fine-tuning dataset size on the ``boost in performance'' (if any) provided by pre-training. Also, we assess the impact on the achieved performance of different combinations of pre-training objectives. Our findings suggest that:

\begin{enumerate}
\item Pre-training is extremely useful when the pre-training dataset is substantially larger than the fine-tuning one, while it does not help when the fine-tuning dataset is of comparable size.

\item The \textsc{mlm} pre-training objective represents a safe choice for all tasks we investigated, being almost always the best-in-class; (iii) specialized pre-training objectives only help if they strictly resemble the fine-tuning task and can provide the model with knowledge that cannot be captured by generic objectives.
\end{enumerate}

We release all code and data used in our study in a comprehensive replication package \cite{replication}.
\section{A SLR on Pre-training Objectives Used to Automate Code-Related Tasks} \label{sec:slr}

\secref{sub:slrDesign} describes the design of our SLR following the guidelines by Kitchenham and Charters \cite{Kitc2007a}. The achieved results are discussed in \secref{sub:slrResults}.  

\subsection{Study Design} \label{sub:slrDesign}

Our SLR aims at answering the following research question: \emph{What are the pre-training objectives used in the SE literature exploiting transformers to automate code-related tasks?}

With ``code-related tasks'' we refer to any task involving source code as input and/or output of the model. For example, code summarization is considered a code-related task, since it takes as input a code component to summarize in natural language. Differently, using transformers to automate ``sentiment analysis'' on software-related artifacts (\eg discussions in issue trackers) is not considered relevant for our SLR. Answering this research question will inform the study presented in \secref{sec:impact}, in which we experiment with representative pre-training objectives from the state-of-the-art.

\subsubsection{Relevant Study Identification}

We used the following digital libraries to search for primary studies: ACM Digital Library \cite{ACM}, IEEE Xplore Digital Library \cite{IEEE}, Springer Link Online Library \cite{Springer}, Wiley Online Library \cite{Wiley}, Elsevier ScienceDirect \cite{Elsevier}, and Scopus \cite{Scopus}. Google Scholar was not considered as an option due to the lack of quality control, clear indexing guidelines, and missing support for data download \cite{Hale2017a}. The following search query has been run on the search engines integrated in each of these online databases:

\begin{flushleft}
\small
\textbf{full text} CONTAINS\\
\hspace{1.1cm}(``\emph{pretrain}'' OR ``\emph{pretrained}'' OR ``\emph{pretraining}'' OR\\
\hspace{1.2cm}``\emph{pre-train}'' OR ``\emph{pre-trained}'' OR ``\emph{pre-training}'' OR\\
\hspace{1.2cm}``\emph{transfer learning}'') AND\\\textbf{publication date} IS FROM \emph{01.01.2007} TO \emph{02.02.2022} AND \textbf{publication venue} CONTAINS\\
\hspace{1.1cm}(``\emph{software}'' OR ``\emph{program}'' OR ``\emph{code}'')
\end{flushleft}

The composition of the query is the result of a trial-and-error procedure performed by the three authors. The query searches for the listed terms (\eg pretrain, pretrained) in the full text of the articles (\ie title, keywords, abstract, main text, references). The date interval has been defined by conservatively collecting papers starting from 2007, year in which we found a first mention to  the notion of ``transfer learning'' in a SE-related article \cite{Waad:2007}, and using the date in which the search has been performed as the end of the interval (02.02.2022). 

Finally, based on the authors' knowledge of the existing SE publication venues, we only searched for articles published in venues containing at least one of three keywords: software, program, and code. We acknowledge that there might be relevant articles published in related fields (\eg artificial intelligence) that our query would exclude. However, our focus was indeed on the SE research community and these keywords should capture most of the relevant venues. 

Some of the search engines (\ie Springer, Wiley, Elsevier, and Scopus) allow to specify a \emph{discipline} of interest. Such a feature is useful to limit the retrieved false positive instances. 

In all online libraries we selected ``Computer Science'' as discipline. In addition, Springer also allows to specify sub-disciplines, for which we selected ``Software Engineering/Programming'' and ``Operating Systems''. While the latter might not be fully relevant, we decided to include it to be more conservative. Links with the exact queries we have run are publicly available \cite{replication}.

\begin{table}[ht]
\centering
    \caption{Articles returned by the queried digital libraries}
    \label{tab:search}
    {\footnotesize
    \begin{tabular}{@{}lr@{}}
    \toprule
    \textbf{Source}                       & \textbf{Returned Articles} \\ \midrule
    ACM Digital Library          & 623                \\
    IEEE Xplore Digital Library  & 850                \\
    Springer Link Online Library & 1,167                 \\
    Wiley Online Library         & 57                 \\
    Elsevier ScienceDirect       & 288                 \\
    Scopus                       & 1,139                \\ \midrule
    Total (including duplicates)                       & 4,124                \\ \midrule
    \textbf{Total (excluding duplicates)}                       & \textbf{2,343}                \\ \bottomrule
    \end{tabular}
    }
\end{table}

\begin{table*}[ht]
	\centering
    \caption{Inclusion and exclusion criteria}
    \label{tab:criteria}
    {\footnotesize
    \begin{tabular}{@{}lp{17cm}@{}}
    \toprule
    & {\bf Inclusion Criteria}\\
    \midrule
    IC1  & The paper must be peer-reviewed, published at SE conferences, workshops, or journals. Such a criterion is particularly important in the snowballing phase described later, in which we ignore all referenced preprints (\eg those published on arXiv.org).
    \\
    IC2  & The PDF of the paper must be available online. If the PDF was not available in the online library of interest, we tried to search it on Google. 
    \\
    IC3 & The paper must present and/or evaluate technique(s) to automate a code-related task.                \\
    IC4 & The proposed/experimented technique(s) must be built on top of a pre-trained transformers model. The pre-training of the model can either have been done directly in the paper or the authors may have used an already existing pre-trained model.           \\ 
    \midrule
    & {\bf Exclusion Criteria}\\
    \midrule
        EC1  & The paper is not written in English.
    \\
    EC2 & The paper has been published in a conference/workshop and later on extended to a journal. We only keep the journal paper to avoid redundancy. 
    \\
    EC3 & The paper is not a full research publication (\eg doctoral symposium articles, posters, ERA track). We exclude all papers having less than six pages. The rationale for such a filter is to remove papers that may not have been subject to the same peer-review process typical of full research papers.
    \\ 
    EC4 & It is unclear from the paper what the adopted pre-training objective is. Such information is instrumental for the goal of our SLR.
    \\ \bottomrule
    \end{tabular}
    }
\end{table*}

\begin{table}[ht]
	\centering
    \caption{Data extraction questionnaire}
    \label{tab:data}
    {\footnotesize
    \begin{tabular}{@{}lp{7cm}@{}}
    \toprule
    {\bf No.} & {\bf Question}\\
    \midrule
    1 & Which code-related task has been automated?\\
    2 & Which specific transformer-based model has been used?\\
    3 & Has the model been pre-training in the paper?\\
    4 & If ``no'' to question 3: Which already pre-trained model has been exploited by the authors?\\
    5 & Which pre-training objectives have been used?\\ \bottomrule
    \end{tabular}
    }
\end{table}

\begin{table*}[ht]
	\centering
    \caption{Pre-training objectives identified in the SLR}
    \label{tab:objectives}
    {\scriptsize
    \begin{tabular}{@{}llp{10cm}p{2.5cm}@{}}
    \toprule
    {\bf Acronym} & {\bf Name} & {\bf Description} & {\bf References}\\\midrule 
    
    \textsc{mlm} & Masked Language Model & Masks $X$\% of tokens (usually 15\%) in the instance (\eg a function) and asks the model to guess the masked tokens based on their bidirectional context. The model knows how many tokens have been masked, since each of them is replaced with a special token (\eg $<$MASK$>$). & \cite{ahmed2021learning, Svyatkovskiy:fse2020, zafar2019towards, chakraborty2021multi, zhou2021finding, he2020structure, qu2021leveraging, gao2021automating, liu2021learning, mariani2021semantic, ghadhab2021augmenting, ciniselli2021empirical, Mastropaolo:icse2021, lin2021traceability, mastropaolo2021empirical, sun2021task, zhou2021smartgift, Zhou:icsme2021, guo2020graphcodebert, ahmad2021unified}\\\midrule
    
    \textsc{nsp} & Next Sentence Prediction & Given two sentences (or two statements) asks the model to guess whether they follow each other. & \cite{zafar2019towards,henkel2021shipwright,gao2021automating,liu2021learning,mariani2021semantic,lin2021traceability,sun2021task,zhou2021smartgift}\\\midrule
    
    \textsc{ulm} & Unidirectional Language Model & A left-to-right language modeling task, asking the model to guess one masked token in an instance by only considering the leftward tokens (\ie the tokens preceding the masked one). & \cite{liu2020multi, heyman2021natural, jiang2021cure}\\\midrule
    
     \textsc{ti} & Token Infilling & Masks a random number of contiguous tokens and asks the model to predict them. Differently from MLM, TI does not suggest to the model how many tokens have been masked, since the sequence of masked tokens is replaced with a single special token (\eg $<$MASK$>$). & \cite{chakraborty2021multi}\\\midrule
    
    \textsc{td} & Token Deletion & Deletes random tokens from the instance expecting the model to reintroduce them where needed. TD is similar to MLM, but without suggesting the model where tokens have been masked. & \cite{chakraborty2021multi}\\\midrule
    
    \textsc{rtd} & Replaced Tokens Detection & Replaces random tokens in the instance with other tokens. The model must guess which are the non-original tokens (\ie those that have been replaced). & \cite{zhou2021finding}\\\midrule

    \textsc{so} & Sentence Ordering & Given two sentences (or two statements) asks the model to guess whether they order. & \cite{asyrofi2020crossasr}\\\midrule
    
    \textsc{im} \faCode & Identifiers Masking & Masks the identifiers in the code instance and asks the model to guess the masked identifiers. & \cite{liu2020multi}\\\midrule
    
    \textsc{plc} \faCode & Prog. Language Classification & Given a sequence of code tokens asks the model to identify its programming language. & \cite{Svyatkovskiy:fse2020}\\\midrule
        
    \textsc{gsm} \faCode & Generative State Modeling & Given assembly code and a small subset of its execution states (\eg register values), asks the model to reconstruct the complete set of its execution states. & \cite{pei2021stateformer}\\\midrule
    
    \textsc{ep} \faCode & Edge Prediction & Masks edges in data-flow graph belonging to 20\% of nodes randomly selected and asks the model to predict them. & \cite{guo2020graphcodebert}\\\midrule
    
    \textsc{na} \faCode & Node Alignment & Similar to data flow EP. Instead of predicting edges between nodes, the model is asked to predict edges between code tokens and nodes. Such a task is performed to align the source code-data flow representations. & \cite{guo2020graphcodebert}\\\midrule
    
    \textsc{cs} \faCode & Code Summarization & Provides as input to the model a function and asks to summarize it in natural language. & \cite{wei2019code}\\
    
    \bottomrule
    \end{tabular}
    }
\end{table*}

\tabref{tab:search} reports the number of articles returned from each digital library (complete list in \cite{replication}). Overall, 4,124 articles were returned, which were reduced to 2,343 by excluding duplicates. Given the very high number of articles, we decided to perform a further cleaning step before starting looking into the papers. We extracted the set of 302 venues in which the articles have been published and two of the authors independently validated them deciding whether to include or exclude them. We excluded venues unrelated to SE or not being international conferences/journals. An open discussion was performed to reach an agreement on the 53 cases of conflict (17\%). As output of this process we kept 163 publication venues as valid, excluding 1,407 papers published in the excluded venues. Examples of excluded publication venues are ``\emph{Computer Methods and Programs in Biomedicine}'' and the ``\emph{Brazilian Symposium on Programming Languages}''. As output we obtained 936 candidate primary studies. \smallskip

\textbf{Study Filtering.} The 936 papers were equally distributed among the three authors. Each author was in charge of inspecting  the paper and decide whether to include or exclude it. Inclusion and exclusion criteria are listed in \tabref{tab:criteria}. As a guideline, the authors agreed on including the paper in case of doubt, since a double-check was foreseen in the study filtering process. Indeed, despite the availability of the selection criteria as reference, such a process still remains highly subjective. A total of 77 papers survived this first analysis. Then, to at least partially address the subjectivity issue, we applied the following procedure. First, we randomly selected 30 papers excluded by each author, asking one of the other two authors to double check whether the papers were actually to be excluded. For all 90 randomly selected papers (30 $\times$ 3 authors), no conflicts arisen, showing consistency in the exclusion criteria applied by the authors. The papers included by each author were also all double-checked by one of the other two authors. Out of the 77 papers included in the first round, 30 made it into the final list of papers, including a SLR that we kept as secondary study for the subsequent snowballing step. 

Cases of disagreement have been discussed among all authors to reach consensus. Note that the decrease brought by the double-check we performed (77 $\rightarrow$ 30) was expected, considering that in the first pass on the papers we decided to be inclusive in case of doubts.

\textbf{Backward Snowballing.} The included papers were split among the authors, with each of them in charge of reading the reference list and identify possible relevant papers. At this stage we relaxed one of our inclusion criteria (IC1): We agreed to include papers published in venues outside of SE as long as they were presenting pre-trained models that have then been exploited to automate code-related tasks in publications appeared in SE venues. Eight papers were added through snowballing. Also in this case, a second author double-checked each of them and, through open discussion among all authors, we finally agreed to include four of them as relevant. This led to the final list of \relevant primary studies included (30 - 1 secondary study + 4 output of the snowballing).

\subsubsection{Data Extraction and Analysis}
The data extraction was performed following the questionnaire in \tabref{tab:data}. While most of the questions are self-explanatory, it is worthwhile to clarify 3 and 4. Our focus is only on papers using pre-trained transformers to automate code-related tasks. However, the pre-training could have been done by the authors of the papers or being the result of a pre-trained model made available in previous work (question 3). If the authors reused an already pre-trained model, then by answering question 4 we expect to know from which reference the pre-trained model has been taken. Question 5 is the ultimate goal of our SLR, which will inform our subsequent experiments.

\subsection{Results Discussion} \label{sub:slrResults}
\tabref{tab:objectives} lists the pre-training objectives we identified. Each pre-training objective is identified by an acronym we will use to refer to it in the text. If the acronym has a \faCode~icon close to it, this indicates that the pre-training task is specific for code, otherwise the pre-training objective is ``generic'' and can be applied to any sort of data that can be fed to the model as a stream of tokens. For each pre-training objective \tabref{tab:objectives} also reports a short description and references to primary studies in the SLR having a pre-trained model using it.

\eject

Without surprise, the most used pre-training objectives are those that the SE community inherited from the NLP community, such as the classic \textsc{mlm}, randomly masking $X$\% of tokens in an instance that, in the case of SE research, could be for example a code function. \textsc{mlm} is used in 21 of the papers included in our SLR. Variations of this pre-training objective are \textsc{ti}, \textsc{ulm}, \textsc{td}, and \textsc{rtd} (see \tabref{tab:objectives} for their description) that are, however, less popular in SE. Among them, the \emph{Replaced Token Detection} (\textsc{rtd}) objective has been proposed in the paper introducing the pre-trained CodeBERT model \cite{feng2020codebert}. CodeBERT is gaining substantial popularity in the SE literature especially when it comes to papers published in 2022 that, due to the time in which we run the search query, are not included in our SLR.

While the above-described objectives work at token-level granularity, \emph{Next Sentence Prediction} (\textsc{nsp}) and \emph{Sentence Ordering} (\textsc{SO}) aim at providing the model with knowledge related to sentence-level relationships. In SE, both can be used for example to ``teach'' the model the correct order of code statements in a given function. \textsc{nsp} is the second most-popular pre-training objective in our set of papers, with 9 articles exploiting it. The popularity of \textsc{nsp} is mostly due to the adoption of the BERT pre-trained model \cite{bert} in SE.

The bottom of \tabref{tab:objectives} lists the objectives specifically designed for code. Despite being pre-training objectives, some of them are targeting a specific task, like for example \emph{Code Summarization} (\textsc{cs}). The latter has been instantiated in \cite{Yang:icpc2021} as a task in which the model was asked, given a Java method, to generate its textual summary (\ie first Javadoc sentence). 

The final code-related task that the authors wanted to automate after the fine-tuning phase was a natural language description of smart contract. Thus, the \textsc{cs} objective started providing the model with knowledge about the code-to-NL translation task. This is a concrete example of pre-training objective already tailored for the specific downstream task of interest. Overall, as it can be seen from \tabref{tab:objectives}, code-specific pre-training objectives have been usually adopted only in the paper proposing them.

\secref{sec:impact} describes how we use the findings of our SLR to select the pre-training objectives for our experiments.
\section{Studying the Impact of Pre-training} \label{sec:impact}

We aim at answering the following research questions:

\textbf{RQ$_1$:} \textit{To what extent is the effectiveness of pre-training influenced by the size of the fine-tuning dataset?} RQ$_1$ investigates if pre-training is still useful when abundant fine-tuning instances are available (\eg can be automatically generated). 

We assess the impact of pre-training when the model is fine-tuned on a dataset (i) being substantially smaller and (ii) having a size comparable to the pre-training dataset. 

\textbf{RQ$_2$:} \textit{To what extent does the choice of the pre-training objective impact the performance of transformer models?} Our second research question focuses on the impact on the model's performance of the used pre-training objectives, with a particular focus on comparing general \emph{vs} task-specific objectives, also investigating combinations of multiple objectives. \smallskip

The context of our experiments are three code-related tasks for which DL-based solutions have been proposed in the past. For reasons we will explain later, RQ$_1$ focuses on \emph{code summarization} and \emph{code completion}, while RQ$_2$ also includes bug-fixing. All our experiments are run on Java datasets defined at method-level granularity (\ie fixing a bug in a method, summarizing a method in natural language, and complete missing code statements in a method). 

\subsection{Transformer Model} 
As representative of transformers \cite{vaswani2017attention}, we adopt the T5 proposed by Raffel \etal \cite{raffel2019exploring}, that has been already used in SE to automate code-related tasks \cite{Tufano:icse2022,isha:icscc,Berabi:mlr,Mastropaolo:tse2022,Ciniselli:tse2021}. 

Raffel \etal proposed several variants of T5, differing in number of trainable parameters. We use the \emph{small} variant, featuring a total of $\sim$60M parameters resulting from 6 layers in both the encoder and the decoder each having a dimensionality of 512, 8-headed attention, and an output dimensionality of 2,048. While larger T5 versions are likely to achieve better performance, the training cost increases with the number of parameters. Considering the number of models that we need to train in our study (\ie 36 different models), we opted for the smaller T5 version. Indeed, our goal is not to achieve state-of-the-art performance in the automated code-related tasks, but rather to study the impact of pre-training on the model's performance in different circumstances. 

\subsection{Pre-training Objectives} \label{sub:objectives} 
We describe the pre-training objectives used in our RQs. We anticipate that they have been applied on a pre-training dataset (detailed in \secref{sub:pretraining-dataset}) which is composed by 1M pairs of Java $\langle$\emph{method}, \emph{javadoc}$\rangle$. 

Concerning RQ$_1$, we only use the \emph{Masked Language Model} (\textsc{mlm}) objective, the one mostly used in the literature. Indeed, the focus of RQ$_1$ is not on the impact of different pre-training objectives (RQ$_2$), but rather on the boost provided by pre-training when the fine-tuning dataset is substantially smaller than the pre-training one or, instead, has a comparable size.

For RQ$_2$, based on the findings of our SLR, we selected three generic pre-training objectives to experiment with. 

First, we picked the two currently being the most popular in SE, namely the \textsc{mlm} (with 15\% of masked tokens) and the \emph{Next Sentence Prediction} (\textsc{nsp}). As third ``generic'' objective, we selected the \emph{Replaced Tokens Detection} (\textsc{rtd}) that, as previously explained, is gaining popularity since used in CodeBERT \cite{feng2020codebert}, adopted in several recent studies (see \eg \cite{Mashhadi:msr2021,Zhang:ist2022,Pan:appsci2021}). While \textsc{mlm} and \textsc{nsp} are straightforward to understand, a clarification is needed on \textsc{rtd}. The latter starts by randomly selecting 15\% of tokens to be replaced. However, the replacements for these tokens are not randomly selected from a vocabulary, but picked based on the recommendation of an $n$-gram model ($n=3$) trained on the pre-training dataset. An $n$-gram model can predict a single token likely to follow the $n-1$ tokens preceding it. As suggested in \cite{feng2020codebert}, we used it to identify suitable alternatives for the tokens to be replaced, thus making the pre-training task (\ie identify which tokens in an instance have been replaced) more challenging. Given a token to replace $T_i$, we run the $n$-gram model by providing it as input with the two tokens preceding it (T$_{i-2}$, T$_{i-1}$) and collect the ranked list of candidate tokens that, accordingly to the $n$-gram model, is likely to follow T$_{i-2}$ and T$_{i-1}$. 

The ranked list features on top the most likely token $T_c$: If $T_c$ is different from the token to replace $T_i$, we use $T_c$ for the replacement. Otherwise, we take the token in second position.

On top of the three generic objectives, we also experiment with three pre-training objectives tailored for the downstream tasks at hand. For \textbf{bug-fixing}, we pre-train the model through the \emph{Injected-Mutants Fixing} (\textsc{imf}) objective. 

The idea is to mutate each method $M$ in the pre-training dataset by injecting artificial bugs in it, creating a mutant $M_m$. During pre-training the \textsc{imf} objective provides T5 with $M_m$ as input and asks it to generate $M$ (\ie to fix the bug). One challenge we faced was the selection of the mutation testing framework to use. We considered tools such as $\mu$\emph{Java} \cite{mujava}, \emph{PIT} \cite{pit}, \emph{javaLanche} \cite{javaLanche}, and  \emph{Jester} \cite{Jester}. PIT was the only one supporting recent versions of Java. However, since it works at Byte code level, PIT requires the input code to mutate to be compilable. This is problematic in our context since the 1M methods in our pre-training dataset come from thousands of software projects, several of which are likely to be unbuildable \cite{Tufano:jsep}. For this reason, we built a source code-level mutation tool using Javaparser \cite{javaparser}. Our tool implements the 11 mutation operators belonging to the ``default group'' in PIT \cite{operators} (\ie \emph{invert negatives}, \emph{empty returns}, \etc). Given a Java method $M$, our tool builds its AST and, using it, identifies the set of mutation operators that can be applied on $M$. For example, the \emph{empty returns} operator replaces return values with ``empty'' values (\eg \texttt{""} if $M$ returns a \texttt{String}, \texttt{Collections.emnptyList()} if $M$ returns a \texttt{Collection}, \etc), and can only be applied to methods returning a value. 

Finally, assuming that $n$ operators can be applied to $M$, $n$ mutants (\ie $n$ versions of $M$) are generated, each implementing one of the applicable operators.

Concerning the \textbf{code summarization} task, we consider the \emph{Method Name Generation} (\textsc{mng}) as a tailored pre-training objective. During pre-training, T5 takes as input a Java method and it is required to synthesize an appropriate name for it, based on the idea that the method name represents an \emph{extreme summary} of the method \cite{allamanis2016convolutional}.

Finally, concerning \textbf{code completion}, we focus on the challenging task of predicting entire code blocks, as recently attempted by Ciniselli \etal \cite{Ciniselli:tse2021}. A code block is defined as the code enclosed between two curly brackets (\eg the code executed when an \texttt{if}/\texttt{else}/\texttt{else if} condition is satisfied). To prepare the model for such a downstream task, we devised \emph{Code Block Selection} (\textsc{CBS}) as a tailored pre-training objective. Given a Java method in the pre-training dataset, we randomly mask a code block in it, and ask the model to decide which of two candidate code blocks is the correct one to complete the method. This pre-training is expected to prepare the model for the more challenging downstream task of generating masked code blocks from scratch.

\subsection{Pre-training Dataset} \label{sub:pretraining-dataset}
The same pre-training dataset is used across both RQs. To build it, we used the GitHub search tool by Dabi\'c \etal \cite{Dabic:msr2021} to identify GitHub Java projects having at least 5 contributors, 50 commits, 10 stars and not being forks of other projects. These selection criteria aimed at removing personal projects from our selection. We ended up with 14,645 valid repositories, that we parsed to extract 64,546,432 Java methods. 

\eject

Since among the downstream tasks we experiment with there is \emph{code summarization}, we wanted to make sure that each pre-training instance was composed by both source code (instrumental for all three tasks) and natural language (useful for \emph{code summarization}). Also, recent studies \cite{Tufano:testGen} showed that pre-training models on both natural language and code (as opposed to code only) is beneficial when dealing with code-related tasks. For these reasons, methods without Javadoc have been excluded, leading to 17,758,579 $\langle$\emph{method}, \emph{javadoc}$\rangle$ pairs. 

We then started processing our dataset to clean it and remove problematic instances. We excluded all pairs meeting one of the following conditions: (i) the Javadoc, while present, is an empty string; (ii) the method has an empty body; (iii) the method is annotated with \texttt{@Test}; (iv) the method does not end with a \texttt{\}} (this may happen in case of parsing errors when we extract the methods). We excluded test methods since none of our tasks is test-related, and we preferred to create a more cohesive pre-training dataset featuring only methods from production code. We only consider in our dataset the first part of the Javadoc comment (\ie the one summarizing the method in natural language) excluding the Javadoc tags (\eg \texttt{@param}, \texttt{@author}). Once done with this basic filtering, the first author manually inspected hundreds of instances in the dataset to identify other sources of noise. 

Four main issues were identified: (i) non-English Javadoc comments; (ii) instances containing non-ASCII characters; (iii) comments containing special symbols/tags which may not help with learning textual patterns in Javadoc; and (iv) comments not representing code summaries, but rather notes written by developers (\eg TODOs). We removed all non-English Javadocs using two Python libraries: \emph{langid} \cite{langid} and \emph{cld3} \cite{pycld3}. We keep an instance in the dataset only if both libraries classified the Javadoc as English text. We replaced all non-ASCII math characters with their corresponding ASCII representation (\eg we replaced ``±’’, with ``+-'') and removed all instances featuring non-Latin characters. Additional cleaning aimed at removing from the Javadoc sequences of characters used for formatting (\eg ``\texttt{-----}'') or special markdown tags (\eg we replace \texttt{\{@class ClassName\}} with \texttt{ClassName}). We also replace any embedded link in the Java method and/or in the Javadoc with a special tag ``\texttt{<LINK\_i>}", with \texttt{$i$} being an integer ranging from 0 to $n-1$, where $n$ is the number of links in the instance. If the same link is found both in the method and in its Javadoc, they are replaced with the same special tag with the same index. 
Finally, since the collected methods came from different projects possibly using different coding styles, we formatted all instances using the Javaparser \cite{javaparser} library.  We also performed additional (minor) cleaning steps that we do not document here due to lack of space. However, we publicly release our cleaning script as part of our replication package \cite{replication}.

After this process, we removed all instances longer than 512 tokens (\ie the number of tokens used to represent both the method and its Javadoc was higher than 512), as also done by previous work using DL to automate code-related tasks (see \eg \cite{leclair-mcmillan-2019-recommendations,Tufano:testGen,Chen:2019}): 4,821,922 instances were left.

As a last step, we excluded instances that are not suitable for one or more of our pre-training objectives. The two objectives which are not applicable to all possible instances are the \emph{Injected Mutants Fixing} (\textsc{imf}) and the \emph{Code Block Selection} (\textsc{cbs}): We removed (i) 219,863 instances for which none of the 11 mutation operators we support can be applied; and (ii) 2,994,723 which did not have any code block to mask. 

From the set of remaining instances, we randomly pick 1M of them to create our pre-training dataset. While, in theory, all $\sim$1.6M  remaining instances were valid, we capped the size of the pre-training dataset to limit the time needed to perform several training epochs.

\subsection{Fine-tuning Datasets} A different fine-tuning dataset has been built for each of the three subject downstream tasks. Due to its limited size, the bug-fixing dataset has only be used in the context of RQ$_2$, since it was not possible to create a version of it large enough (\ie having a size comparable to our pre-training dataset) to answer RQ$_1$ as well. 

\subsubsection{Bug-fixing} We exploit the dataset used by Chen \etal \cite{Chen:tse2021} when presenting SequenceR, a sequence-to-sequence model trained on 35,578 one-line Java bug fixes (\ie commits fixing a bug by only changing a single line of code). 

The training set consists of pairs featuring the buggy code and the corresponding fixed code, and it is accompanied by a validation and a testing set featuring additional 4,711 one-line bug fixes each. The buggy code includes a ``buggy line'' explicitly marked with two special tokens (\ie $\langle$START\_BUG$\rangle$ and $\langle$END\_BUG$\rangle$) and being part of a ``buggy method''. In addition to that, the buggy code also includes contextual information extracted from the ``buggy class'' (\eg its constructor). The fixed code the model is expected to generate includes, instead, only the ``fixed line'' (\ie revised version of the ``buggy line''). 

Before using this dataset, we pre-processed it to make it more ``aligned'' to our pre-training dataset, and in particular to the tailored pre-training objective we devised for the \emph{bug-fixing} task (\ie \emph{Injected-Mutants Fixing}). First, our pre-training dataset only features Java methods with its related Javadoc, excluding any class-related information. Similarly, the \textsc{imf} objective provides as input to the model a mutated Java method without any additional contextual information nor special tag signaling the injected bug. Thus, we processed the buggy code of the dataset by Chen \etal \cite{Chen:tse2021} to only include the buggy method without the special tokens marking the buggy line. Second, as done for the pre-training instances, we formatted the code using Javaparser \cite{javaparser}, to have a coherent code representation. Finally, we removed any duplicated method already present in our pre-training dataset. This process left us with 25,901 instances that we split into training (80\%), validation (10\%) and test (10\%) set. 

\subsubsection{Code summarization} We use the FunCom dataset \cite{leclair-mcmillan-2019-recommendations,funcom}, featuring 2,149,120 instances, with each of them being composed by a Java method and its associated Javadoc comment. FunCom has been curated to only include English comments and exclude auto-generated files. 

We start from the ``Filtered dataset" version \cite{funcom} consisting of not processed instances. We perform on them the same cleaning process used for the pre-training dataset (\eg removing instances containing non-Latin characters) and remove any duplicate with between FunCom and our pre-training dataset. From the  1,898,437 instances left, we create two fine-tuning datasets needed to answer RQ$_1$. For the first (\emph{large-ft}), we randomly select 1M instances, splitting them into training (80\%), validation (10\%), and test (10\%). This fine-tuning dataset will be used in RQ$_1$ as representative of a fine-tuning dataset having a size of the same order of magnitude of the pre-training dataset. For the second (\emph{small-ft}), we randomly select 25,901 instances, the same number of instances in our \emph{bug-fixing} fine-tuning dataset. The idea is indeed to create a second fine-tuning dataset being substantially smaller than the pre-training one, as it usually happens when working on tasks characterized by a scarcity of training data. Also \emph{small-ft} followed the usual training (80\%), validation (10\%), and test (10\%) split. Both datasets are used to answer RQ$_1$, while only \emph{small-ft} is used in RQ$_2$. Indeed, as our RQ$_1$'s findings will show, pre-training is mostly useful when a small fine-tuning dataset is available. Thus, we experiment the impact on performance of the pre-training objectives (\ie RQ$_2$) when using the \emph{small-ft} dataset.

\subsubsection{Block-level code completion} Following Ciniselli \etal \cite{Ciniselli:tse2021}, we aim at building a fine-tuning dataset in which Java methods having a masked block of up to three statements are provided as input to T5, which is in charge of generating the masked block. We start from the 1,569,889 Java methods in the CodeSearchNet dataset \cite{husain2019codesearchnet}. We applied a cleaning process similar to the one described for the pre-training dataset (\eg checking that the method body is not empty, that the method is not a test method, \etc), removed methods not containing at least one code block composed by at most three code statements (2,847). Then, we followed the  training procedure by Ciniselli \etal \cite{Ciniselli:tse2021}: Given $k$ the number of blocks identified in a method $M$, we create $k$ versions of $M$, each one having a specific code block masked. As previously said, only blocks composed by at most three statements are masked. We then remove instances longer than 512 tokens (333,955). Such a process resulted in a dataset composed by 1,823,977 instances. Finally, similarly to what explained for code summarization, we create two versions of the \emph{code completion} fine-tuning dataset: \emph{large-ft} featuring a total of 1M randomly selected instances, and \emph{small-ft} featuring 25,901 instances.\smallskip

\begin{table}[t]
\centering
    \caption{Pre-training and fine-tuning datasets used in our study}
    \label{tab:datasets}
    {\footnotesize
    \resizebox{\linewidth}{!}{
    \begin{tabular}{lrrrcc}
    \toprule
    \textbf{Dataset}                       & \textbf{Training}  & \textbf{Evaluation} & \textbf{Test} & \textbf{RQ$_1$} & \textbf{RQ$_2$}\\ \midrule
    \textbf{Pre-training}          &                1,000,000 & - & - & \cmark & \cmark\\\midrule
    \textbf{Fine-tuning} & \\
    \hspace{3mm} \emph{Bug-fixing} & 22,321 & 2,790 & 2,790 & \xmark & \cmark\\
    \hspace{3mm} \emph{Code summarization} & \\
    \hspace{6mm} \emph{large-ft} & 800,000 & 100,000 & 100,000 & \cmark & \xmark\\
    \hspace{6mm} \emph{small-ft} & 22,321 & 2,790 & 2,790 & \cmark & \cmark\\
    \hspace{3mm} \emph{Code completion} & \\
    \hspace{6mm} \emph{large-ft} & 800,000 & 100,000 & 100,000 & \cmark & \xmark\\
    \hspace{6mm} \emph{small-ft} & 22,321 & 2,790 & 2,790 & \cmark & \cmark\\\bottomrule
    \end{tabular}
    }
    }
\end{table}

\tabref{tab:datasets} summarizes the pre-training/fine-tuning datasets, also indicating which dataset has been used in each RQ.

\subsection{Experimental Procedure} \label{sub:procedure}
The training of the models has been performed using a 2x2 TPU topology (8 cores) from Google Colab with a batch size of 128 and the \emph{Inverse Square Root} learning rate. 

\subsubsection{Answering RQ$_1$}. We start by fine-tuning (without pre-training) four models, two for the \emph{code summarization} and two for the \emph{code completion} task. 

The models trained within each task differ for the fine-tuning dataset used, being either the \emph{large-ft} (800k training instances) or the \emph{small-ft} ($\sim$22.3k). The fine-tuning has been performed using an early-stopping training strategy by exploiting the evaluation set. In particular, we saved a checkpoint of the model every epoch computing its performance in terms of correct predictions on the evaluation set and stopped the training if the performance of the model did not increase for three consecutive checkpoints (to avoid overfitting). With ``correct predictions'' we refer to cases in which the generated prediction is identical to the target. For code summarization, a correct prediction implies that the summary generated by the model is equal to the one written by developers. In the case code completion, the predicted code block matches the one we masked.

Then, we pre-train a T5 model for 40 epochs on the 1M instances featured in the pre-training dataset using the \textsc{mlm} objective. We fine-tune four versions of it, again two for each task (\ie \emph{code summarization} and \emph{code completion}) differing for the used fine-tuning dataset (\emph{large-ft} or \emph{small-ft}). We used the same early-stopping procedure described above.

This process resulted in an overall of eight models, four being pre-trained and four not being pre-trained. These eight models have been run on the corresponding test sets collecting their predictions. This allows to compare the pre-trained and the not pre-trained models both when using a fine-tuning dataset having a size comparable to that of the pre-training (\emph{large-ft}) or being substantially smaller than it (\emph{small-ft}).

\subsubsection{Answering RQ$_2$}. We experiment with four possible pre-training objectives (and their combinations) for each of the three tasks subject of our study: three ``generic'' pre-training objectives (\ie \textsc{mlm}, \textsc{nsp}, and \textsc{rtd}) that can be applied to any task and the pre-training objective specifically tailored for the given task (\eg \emph{Method Name Generation} for code summarization). Thus, for each task, we start by pre-training and fine-tuning four T5 models, each one using a specific pre-training objective. To avoid confounding factors, we made sure that all pre-trainings (i) exploit exactly the same pre-training instances (\ie the 1M instances in \tabref{tab:datasets}), and (ii) are run for 40 training epochs. Concerning the fine-tuning, we adopt the same early-stopping training strategy described for RQ$_1$. 

The above-described process results in 12 different models being pre-trained and fine-tuned (4 $\times$ 3 tasks). 

\eject

For each task, we evaluate the four fine-tuned models on the corresponding test set (see \tabref{tab:datasets}) in terms of correct predictions, identifying the best performing pre-training objective $pt_o$. The latter has then been combined in pairs with the remaining three objectives. For example, assuming that \emph{Method Name Generation} (\textsc{mng}) results the best pre-training objective among the four experimented for the \emph{code summarization} task, we create three pairs of pre-training objectives including $<$\textsc{mng}, \textsc{mlm}$>$, $<$\textsc{mng}, \textsc{nsp}$>$, and $<$\textsc{mng}, \textsc{rtd}$>$. This provides us with additional three models pre-trained and fine-tuned for each task (9 models overall --- 3 $\times$ 3 tasks). Again, each of these models has then be evaluated on the corresponding test set, identifying the best performing ``pair'' of pre-training objectives for each task. The latter has been used to generate two triplets of pre-training objectives by combining it with the remaining two objectives. For example, assuming $<$\textsc{mng}, \textsc{nsp}$>$ to be the best pair for \emph{code summarization}, we create $<$\textsc{mng}, \textsc{nsp}, \textsc{mlm}$>$, and $<$\textsc{mng}, \textsc{nsp}, \textsc{rtd}$>$. This results in the training and testing of two additional models for each task (6 overall). Finally, we test for each task the full combination of the four corresponding pre-training objectives, thus training one additional model for each task (3 overall). In total, we pre-trained and fine-tuned 12+9+6+3=30 T5 models in RQ$_2$.

The output of such a process is, for each task, the set of predictions generated by the 10 models trained for it (4 using a single pre-training objective, 3 using pairs of objectives, 2 using triplets, and 1 using all four objectives).
 
\subsection{Data Analysis}
Using the generated predictions, in both our RQs we assess the performance of the models by computing the percentage of correct predictions generated (\ie predicted output identical to the expected one). We statistically compare the results achieved by the different models for each task using the McNemar's test \cite{mcnemar}, which is a proportion test suitable to pairwise compare dichotomous results of two different treatments. To account for running multiple test instances (\eg in RQ$_2$ comparing the results of the model pre-trained with \textsc{mlm} with those pre-trained using \textsc{rtd}, \textsc{nsp}, \etc), we adjust \emph{p}-values using the Holm's correction \cite{Holm1979a}. We complement the McNemar's test with the Odds Ratio (OR) effect size. 



\begin{table}[ht]
	\centering
	\caption{RQ$_1$: Impact of pre-training when the fine-tuning dataset size ($|$FT$|$) is $<<$ or $\sim$ to the pre-training dataset size ($|$PT$|$)}
	\scriptsize
	\begin{tabular}{llrr}
		\toprule
		{\bf $|$FT$|$ \emph{vs} $|$PT$|$} & {\bf Task} & {\bf Non Pre-trained} & {\bf Pre-trained (MLM)}\\\midrule
		\multirow{2}{*}{$<<$} & Code Summarization & 1.94\% & 4.73\%\\
		& Code Completion & 2.37\% & 5.05\% \\\midrule
		
		\multirow{2}{*}{$\sim$} &Code Summarization & 16.60\% & 15.98\%\\
		& Code Completion & 30.41\% & 29.11\%\\\bottomrule

	\end{tabular}%
	\label{tab:rq1}%
\end{table}

\section{Results Discussion} \label{sec:results}

We discuss the achieved results by research question. We highlight the main take-aways of our study using the \faLightbulbO~icon.

\subsection{RQ$_1$: Effectiveness of pre-training when dealing with fine-tuning datasets of different sizes} \tabref{tab:rq1} reports the percentage of correct predictions achieved by the non pre-trained T5 and the T5 pre-trained using the \textsc{mlm} objective. Results are reported for the two tasks involved in RQ$_1$ (\ie \emph{code summarization} and \emph{code completion}) in the scenario in which the fine-tuning dataset is (i) substantially smaller (22.3k $<<$ 1M) than the pre-training dataset (\ie the \emph{small-ft} dataset has been used --- top part of \tabref{tab:rq1}), and (ii) of a size similar (800k $\sim$ 1M) to the pre-training dataset (\ie the \emph{large-ft} dataset has been used --- bottom part of \tabref{tab:rq1}).

\faLightbulbO~When the fine-tuning dataset is small, the pre-training, as expected, helps the learning of the model. For \emph{code summarization}, the boost in terms of perfect predictions goes from 1.94\% up to 4.73\%, resulting in a statistically significant difference ($p$-value $<$ 0.001) with an OR=10.3, indicating ten times higher odds of obtaining a correct prediction from the pre-trained model as compared to the non pre-trained one. Similar findings hold for the \emph{code completion} task, with correct predictions growing from 2.37\% to 5.05\% thanks to the pre-training ($p$-value $<$ 0.001, OR=11.7). 

Moving to the bottom part of \tabref{tab:rq1}, two observations can be made. First, with the fine-tuning dataset being 36 times larger (22.3k \emph{vs} 800k) the performance of the model improves dramatically both for the pre-trained and for the non pre-trained model. This is kind of expected considering the larger amount of data from which the model can learn useful patterns. 

Second, \faLightbulbO~when the fine-tuning dataset size is similar to that of the pre-training dataset, we do not observe any boost provided by pre-training, with performance slightly in favor of the non pre-trained model in both tasks ($p$-value $<$ 0.001, OR=1.2 for \emph{code summarization}, and $p$-value $<$ 0.001, OR=1.3 for \emph{code completion}). While such a result may look surprising, it might be partially explained by the well-known ``catastrophic forgetting'' phenomenon affecting neural networks \cite{Robins:1995}: DL models tend to forget previously learned information once new information is provided. Having a large fine-tuning dataset may lead to ``override'' what the model learned during pre-training, making the latter basically useless.

\begin{figure}[h!]
	\centering
	\includegraphics[width=\linewidth]{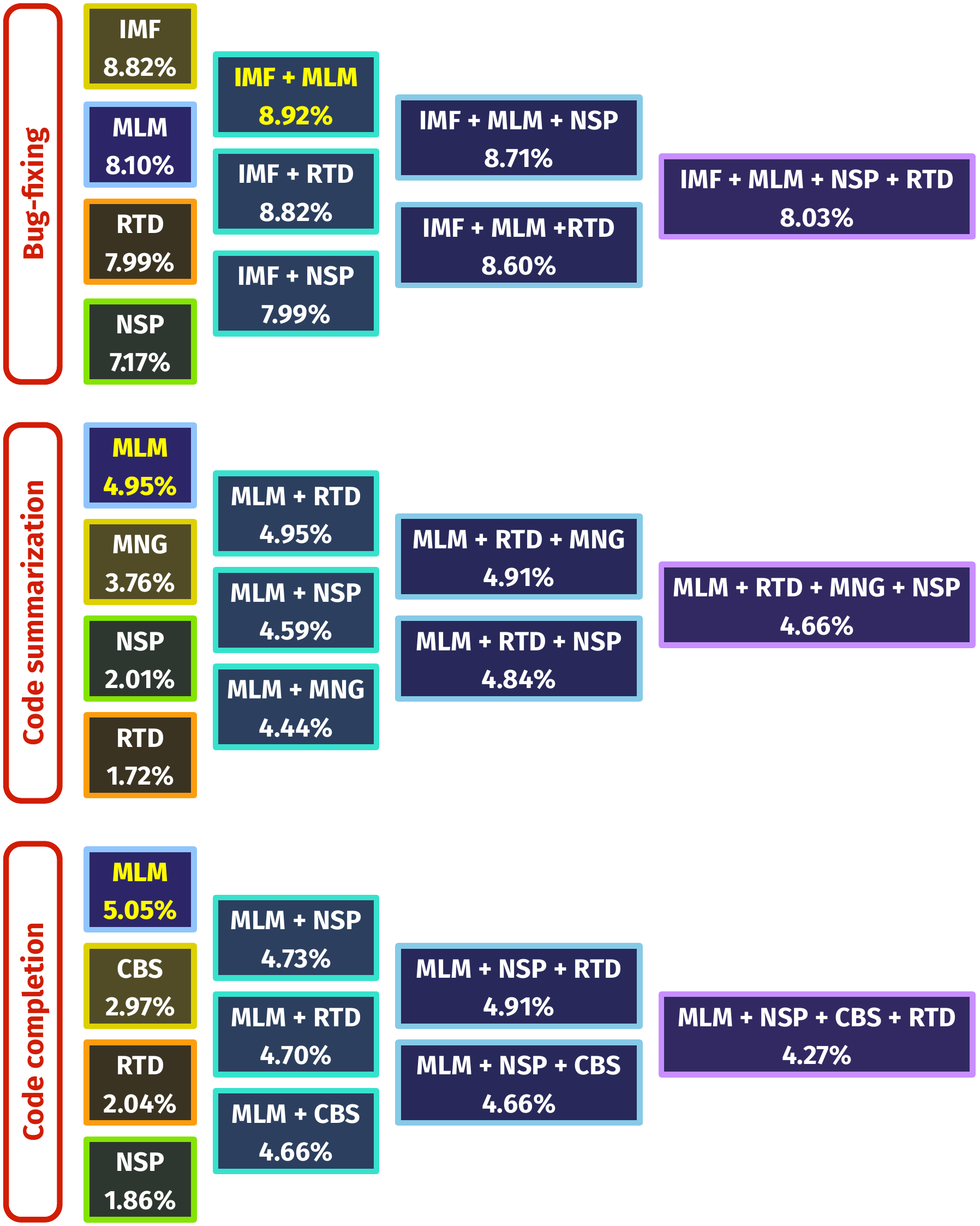}
	\caption{RQ$_2$: Results with different combinations of pre-training objectives}
	\label{fig:resultsRQ2}
\end{figure}

\subsection{RQ$_2$: Impact of pre-training objectives on performance} 
\figref{fig:resultsRQ2} summarizes the results in terms of correct predictions achieved by the 30 models pre-trained using different objectives (10 for each task). These experiments have been run in the scenario in which pre-training helps the most (\ie when using small fine-tuning datasets). From the left to the right of \figref{fig:resultsRQ2} we move from pre-trainings performed with a single objective (\eg \textsc{mlm}) to those involving all objectives relevant for a given task (\eg \textsc{imf} + \textsc{mlm} + \textsc{nsp} + \textsc{rtd}). Within each task and each pre-training group (\ie those involving only one objective, those involving two objectives, \etc), the objectives are sorted from the top to the bottom based on the performance they ensured. 

\eject

For example, for \emph{bug-fixing}, when experimenting with a single pre-training objective, the task-specific objective we devised (\ie \emph{Injected-Mutants Fixing} --- \textsc{imf}) is the best one, followed by \textsc{mlm}, \textsc{rtd}, and \textsc{nsp}. The overall best performing combination of objectives for each task is highlighted with yellow text (\eg \textsc{imf + mlm} for bug-fixing). Given the high number of statistical tests performed (\ie each combination of pre-training objectives has been contrasted against all others, resulting in 45 tests per task), we provide full tables with the adjusted $p$-values and ORs in our replication package \cite{replication}.

Three main lessons can be learned from our results. First, \faLightbulbO~the choice of the pre-training objective can make a substantial difference in the performance of transformers. Within each task, contrasting the best-performing combination with the worst-performing one results in statistically significant differences ($p$-value $<$ 0.001) with ORs of 2.9 (\emph{bug-fixing}), 38 (\emph{code summarization}), and 29 (\emph{code completion}). For example, in the case of \emph{code summarization}, we move from the 4.73\% of correct predictions ensured by \textsc{mlm}, down to the 1.94\% achieved with the \textsc{nsp} objective.

Second, the \faLightbulbO~high effectiveness of the \textsc{mlm} pre-training objective: \textsc{mlm} is involved, alone or in combination with other objectives, in all best configurations we found (\ie \textsc{imf + mlm} for \emph{bug-fixing}, \textsc{mlm} for \emph{code summarization} and \emph{code completion}). 

\eject

Moreover, even in the \emph{bug-fixing} task in which the \textsc{mlm} objective in isolation is not the best-in-class, the difference in performance with respect to the best configuration (\ie \textsc{imf + mlm}) is not statistically significant.

Third, concerning the task-specific objectives we devised, we only observed a (not statistically significant) improvement over \textsc{mlm} for the \emph{bug-fixing} task with the \textsc{imf} objective. The latter provides the model with information substantially different from those that can be captured by \textsc{mlm} and it closely resembles the fine-tuning task. This might not be the case for the other two task-specific objectives. For example, the \emph{Method Name Generation} \textsc{mng} objective devised for \emph{code summarization} is a sort of more specific version of \textsc{mlm}: rather than randomly masking 15\% of tokens in the training instance, the method name is the only masked token the model has to predict. We conclude that \faLightbulbO~task-specific pre-training objectives might boost performance if they (i) capture orthogonal information as compared to non-specific objectives such as \textsc{mlm}; and (ii) strictly simulate the downstream task. However, even in this case, the gain over the classic \textsc{mlm} objective may be limited and should be assessed empirically.
\section{Threats to Validity} \label{sec:threats}

\textbf{Construct validity.} The usage of the correct predictions as metric only provides a limited view about the performance of the experimented models. For example, for \emph{code summarization}, the model might generate a code summary that, while different from the one written by developers, may be semantically equivalent. To partially address this threat, for the \emph{code summarization} task we also computed the BLEU-4 score \cite{Papineni:2002} of the predictions, being the overlap in terms of 4-grams between the predicted and the reference summary. 

Similarly, for the \emph{code completion} task, we compute the normalized token-level Levenshtein distance \cite{levenshtein1966} between the predicted code block and the target one. The latter indicates the percentage of code tokens in the prediction that should be changed to obtain the target code block. The results of this analysis are available in our replication package \cite{replication}. In short, our findings hold by considering these metrics instead of the correct predictions (\eg the best pre-training objective for \emph{code summarization} in RQ$_2$ is the same both by considering the percentage of correct predictions or the BLEU-4).

\textbf{Internal validity.} As a design choice, we did not perform any hyperparameter tuning of T5, using the architecture proposed by the original authors \cite{raffel2019exploring}. However, we compared the different models (\eg with/without pre-training) when using the same exact configuration. Thus, we expect no impact of this choice on our findings.

When we employed the small versions of the fine-tuning datasets, we observed a substantial drop of the models' performance in RQ$_1$. This may pose questions on how realistic is the size of the \emph{small-ft} datasets we used. However, we just mirrored the size we had for the \emph{bug fixing} task, for which the scarcity of training data is a real problem.

\eject

For the SLR, we did not consider papers published in 2022, since we started working on this paper in December 2021 running the search queries in January 2022. We acknowledge that several works relying on pre-trained transformers have been published in the meanwhile. However, the goal of our work was not to be comprehensive in terms of all pre-training objectives ever used in SE, but to get a good overview to inform our main study (\secref{sec:impact}).

Finally, due to the stochastic nature of neural networks, small variations are possible when re-running our trained models on the same test set. This is something we observed in our experiments with a few predictions changing from one run to another. The overall findings are not affected by such minor changes.

\textbf{External validity.} We used T5 as representative of transformers. Other DL models may lead to different results. Also, all our experiments are focused on Java and executed at method-level granularity. Our findings may not generalize to other settings.
\section{Related Work} \label{sec:related}

Given the space constraints, we do not discuss the extensive literature related to the usage of pre-trained models in SE, partially documented in our SLR. We focus on (i) related literature reviews and (ii) empirical studies investigating specific aspects concerning the usage of DL models in SE.

\textbf{Related Literature Reviews.} At the time of writing, we found three SLRs related to the use of DL models in SE. Le~\cite{le2020deep} \etal investigated practices and challenges of using DL for source code modeling and generation, classifying the used architectures, the strategies for dealing with issues such as the large vocabulary of source code, and the targeted tasks. 

Ferreira~\cite{ferreira2021software} \etal conducted a mapping study to understand the usage of DL models in SE. The authors focus the attention on the tasks addressed through DL and the architectures used. 

Watson~\cite{watson2022systematic} \etal also presented a SLR aimed at documenting the use of DL
in SE. The authors summarize 151 manuscripts to answer five research questions, related to (i) the tasks addressed via DL, (ii) the data pre-processing and datasets creation pipeline, (iii) the DL models used, (iv) the reported performance, and (v) the replicability of the studies.

Our SLR has a different goal: We investigated the objectives used to pre-train transformers in the SE literature as a starting point to inform our main study (\secref{sec:impact}).


\textbf{Empirical Studies on DL Models in SE.} 
By considering the non-trivial effort needed to create labeled datasets, Robbes and Janes~\cite{Robbes:icse2019} suggested the use of pre-trained models in SE. They showed that pre-trained models can substantially boost performance when small fine-tuning datasets are available.

Zhou \etal~ \cite{Zhou:icsme2021} investigated the ability of CodeBERT~\cite{feng2020codebert} to generalize beyond its pre-trained data and to reuse embedded knowledge in various SE tasks. The results confirmed the superiority of CodeBERT compared to specialized models. Both pre-training objectives exploited by CodeBERT (\ie \textsc{mlm} and \textsc{rtd}) have been considered in our study.


Paltenghi and Pradel~\cite{Paltenghi:ase2021} studied if the artificial attention mechanisms of DL models behave similarly to skilled  developers comprehending code. The results showed that models and humans tend to focus on different parts of code.


Chen \etal~\cite{Chen:icpc2022} investigated the proposal by Ahmed and Devanbu~\cite{AhmedDevanbu2021} to pre-train DL models on multiple programming languages. The authors reported that multilingual models have worst performance as compared to monolingual ones. 

To the best of our knowledge, our work is the first extensively studying the impact of pre-training on the performance of transformers when automating code-related tasks.
\section{Conclusions} \label{sec:conclusions}

We investigated the impact on the performance of transformers \cite{vaswani2017attention} of the pre-training phase, nowadays adopted in most of the applications of these models to SE tasks. Two aspects have been investigated: (i) the extent to which pre-training helps the learning even when the task at hand allows to build very large datasets for fine-tuning; and (ii) the impact on the model's performance the choice of the pre-training objective(s) can have. 

We found that when the size of the fine-tuning dataset is large enough, approaching that of the pre-training dataset, the pre-training phase is unlikely to help. Instead, it provides a substantial boost of performance for tasks in which the scarcity of training data leads to small fine-tuning datasets. We also observed the major role played by the choice of the pre-training objectives, with different combinations of objectives providing substantially different performance. Pre-training objectives specifically tailored for the downstream tasks can help but, at least in our study, did not result in a significant improvement of performance as compared to the classic \emph{Masked Language Model} task.

\section*{Acknowledgment}
This project has received funding from the European Research Council (ERC) under the European Union's Horizon 2020 research and innovation programme (grant agreement No. 851720). 

\bibliography{main}
\bibliographystyle{IEEEtran}

\end{document}